\documentclass[structabstract]{aa}  
\usepackage{graphicx}
\usepackage{natbib}
\def\msun{M$_{\odot}$}
\def\mdot{$\dot M$}

\begin{document}

\title{Intermediate polars in the {\it Swift}/BAT survey: \\
Spectra and white dwarf masses}

\titlerunning{WD masses of IPs}

   \author{J\"org Brunschweiger
          \inst{1}
          \and
          Jochen Greiner\inst{1}
          \and
          Marco Ajello\inst{1}\thanks{Present address: SLAC/KIPAC, 
              2575 Sand Hill Road, Menlo Park, CA 94025;
              majello@slac.stanford.edu}
          \and 
          Julian Osborne\inst{2}
          }

\authorrunning{Brunschweiger, Greiner, Ajello \& Osborne}

\offprints{jcg@mpe.mpg.de}

   \institute{Max-Planck-Institut f\"ur Extraterrestrische Physik,
              Giessenbachstrasse 1, 85748 Garching, Germany\\
              \email{jcg@mpe.mpg.de}
             \and
             Dept. of Physics and Astronomy,
            University of Leicester, University Road, Leicester LE1 7RH, UK\\
             \email{julo@star.le.ac.uk}
              }
             
   \date{Received 3 November  2008; accepted 9 January 2009}

% \abstract{}{}{}{}{} 
% 5 {} token are mandatory
 
\abstract
{White dwarf masses in cataclysmic variables are difficult to determine 
accurately, but are fundamental for understanding binary system parameters, 
as well as binary evolution.
}
% AIM
{
We investigate the X-ray spectral properties of 
a sample of Intermediate Polars (IP) detected above 15\,keV
to derive the masses of their accreting white dwarfs.
}
% METHODS
{
We use data from the {\em Swift}/BAT instrument which during the first 2.5\,yrs
of operation has detected 22 known intermediate polars. 
The X-ray spectra of these sources
are used to estimate the mass of the white dwarfs.
}
{
We are able to produce a mass estimate for 22 out of 29 of the 
confirmed intermediate
polars. Comparison with previous mass measurements shows good agreement.
For GK Per, we were able to detect spectral changes due to the
changes in the accretion rate.
}
{The {\em Swift}/BAT detector with its combination of sensitivity and all-sky 
coverage provides an ideal tool to determine accurate white dwarf masses in 
intermediate polars. This method should be applied to other magnetic 
white dwarf binaries.}

\maketitle

\section{Introduction}

Cataclysmic variables (CV) are binaries with a white dwarf (WD) accreting 
matter from a late main-sequence star which fills its Roche lobe
\cite[][]{warner95}. Depending on the strength of the magnetic field
of the white dwarf, the matter flowing through the Lagrange point 
either forms a full or partial accretion disk or follows the magnetic 
field lines towards the magnetic poles of the white dwarf.
Intermediate polars (IPs) are a sub-type of cataclysmic variables
(see e.g. the review by \cite{hellier2002}) 
where a magnetic field of order B $\approx$ 1-10 MG \citep{deMartino2004}
disrupts the inner part of the accretion disk. From this disruption radius
the accreting material follows the magnetic field lines towards the 
magnetic poles and finally falls freely onto the surface of the WD,
forming a strong shock near the surface of the WD \cite[][]{aizu73}. 
This so called post shock region  is heated up to a high temperature,
about 10--20\,keV \cite[][]{KingLasota79}.
The main cooling process is thermal Bremsstrahlung. 
The temperature of the post shock region and hence the spectral shape of the 
emitted X-ray radiation depend to first order only on the mass of the WD. 
This makes IPs especially interesting, because their X-ray spectrum
allows a relatively simple estimate of the WD mass 
\citep[see e.g.][]{rothschild81,ishida91,suleimanov05}.

The white dwarf mass is a fundamental parameter in cataclysmic binaries.
It not only governs the dynamics of the orbital motion and the accretion flow, 
but also is a main parameter characterizing the emission of the 
accretion region (see above). Moreover, it plays a fundamental role in 
the binary evolution, and accurate mass distributions of certain 
sub-classes can help in constraining the accretion history.

\begin{table*}
 \caption{IPs detected in the BAT survey  3$\sigma$ in the 
  15--55 keV band ordered by constellation. 
  The first three columns give the BAT position and the offset to
     the optical position.
  The count rate  was averaged over the entire 2.5 yr  observing period.  
In the lower part of the table we report those sources which are not
currently detected in the BAT survey; for those we give the optical 
 position for which the upper limit was computed.
 \label{tbl:werte2}}
 \begin{tabular}{lrrrcrcc}
 \hline 
 \noalign{\smallskip}
   \textbf{Name} & R.A.~ & Decl.~ & Offset & Count rate~~~ & Significance & Exp. time & Sensitivity$^1$\\
  \noalign{\smallskip}
         & (2000.0) & (2000.0) & (arcmin) & (10$^{-5}$  cts/s/det) & ratio & (Ms) & (10$^{-11}$ erg/cm$^2$/s) \\
 \noalign{\smallskip}
 \hline 
 \noalign{\smallskip}
 FO Aquarii          & 334.481 & -8.351 &3.3 &12.4 $\pm$ 0.8  & 16.6& 2.76 & 1.99 \\
 XY Arietis          & 44.038  & 19.441 &4.3 & 5.3 $\pm$ 0.6  & 7.9 & 2.70 & 1.55 \\
 V405 Aurigae        & 89.497  & 53.896 &0.3 & 5.4 $\pm$ 0.6  & 8.7 & 2.70 & 1.66 \\
 MU Camelopardaris   & 96.318  & 73.578 &1.4 & 2.9 $\pm$ 0.5  & 6.7 & 3.46 & 1.60 \\
 V709 Cassiopeiae    & 7.204   & 59.289 &0.7 &16.2 $\pm$ 0.7  & 27.5& 3.65 & 1.61 \\
 V1025 Centauri      & 189.569 & -38.713&4.1 & 1.6 $\pm$ 0.8  & 3.1 & 1.91 & 1.88 \\
 BG Canis Minoris    & 112.871 & 9.940  &2.9 & 5.8 $\pm$ 0.8  & 7.0 & 1.92 & 2.15 \\
 TV Columbae         & 82.356  & -32.818&2.5 &13.0 $\pm$ 0.7  & 20.3& 3.55 & 1.44 \\
 TX Columbae         & 85.834  & -41.032&3.5 & 4.5 $\pm$ 0.5  & 7.8 & 3.81 & 1.31 \\
 V2306 Cygni         & 299.560 & 32.545 &3.4 & 3.1 $\pm$ 0.5  & 4.8 & 3.85 & 1.77 \\
 DO Draconis         & 175.910 & 71.689 &1.8 & 2.7 $\pm$ 0.4  & 7.8 & 4.47 & 1.11 \\
 PQ Geminorum        & 117.822 & 14.740 &2.7 & 6.2 $\pm$ 0.7  & 9.4 & 2.04 & 2.17 \\
 EX Hydrae           & 193.102 & -29.249&3.0 & 5.2 $\pm$ 0.7  & 6.4 & 1.87 & 1.97 \\
 NY Lupi             & 237.061 & -45.478&2.1 &15.2 $\pm$ 0.8  & 15.5& 1.35 & 2.58 \\
 V2400 Ophiuchi      & 258.152 & -24.246&4.2 &10.5 $\pm$ 0.9  & 11.0& 1.30 & 2.52 \\
 GK Persei           & 52.799  & 43.905 &4.4 &18.5 $\pm$ 0.9  & 30.4& 2.88 & 1.58 \\
 AO Piscium          & 343.825 & -3.178 &4.7 & 8.1 $\pm$ 0.6  & 11.1& 3.00 & 1.62 \\
 V1223 Sagittarii    & 283.759 & -31.163&3.0 &19.6 $\pm$ 1.1  & 20.8& 1.31 & 2.54 \\
 V1062 Tauri         & 75.615  & 24.756 &5.3 & 3.1 $\pm$ 0.8  & 5.9 & 2.09 & 2.25\\
 RX J2133.7+5107     & 323.432 & 51.124 &1.9 &10.9 $\pm$ 0.6  & 18.2& 3.99 & 1.42 \\
 Swift J0732.5-1331  & 113.156 & -13.518&2.8 & 6.3 $\pm$ 0.6  & 9.7 & 2.61 & 1.70 \\
 IGR J00234+6141     & 5.740   & 61.686 &5.3 & 2.0 $\pm$ 0.6  & 3.0 & 3.68 & 1.72 \\
 IGR J17303-0601     & 262.590 & -5.993 &0.8 &14.0 $\pm$ 0.8  & 12.7& 1.59 & 2.78\\
 \noalign{\smallskip}
 \hline
 \noalign{\smallskip}
 AE Aquarii          & 310.04 & -0.87 & --&$<0.3$ & 0.4& 2.44 & 2.05\\
 HT Camelopardaris   & 119.26 & 63.10 & --&$<0.1$ & 0.3 & 3.52 & 1.34 \\
 DW Cancri           & 119.72 & 16.28 & --&$<1.5$ & 1.5 & 2.09 & 2.11 \\
 UU Columbae         & 78.05  & -32.69& --&$<0.7$ & 1.7 & 3.55 & 1.50 \\
 DQ Herculis         & 271.88 & -45.86& --&$<1.0$ & 2.5 & 3.75 & 1.56 \\
 WX Pyxidis          & 128.27 & -22.81& --&$<1.4$ & 2.8 & 3.16 & 1.56\\
 \noalign{\smallskip}
 \hline 
 \noalign{\smallskip}
 \noalign{1.  Minimum flux to detect a source at 3$\sigma$ at a given position
	in the 15-55\,keV band.
   }
 \end{tabular}
\end{table*}

Previous attempts to determine WD masses of IPs using X-ray measurements
 have been controversial, since data from different X-ray instruments
led to different mass estimates (\cite{ishida91} with Ginga data, or 
\cite{ramsay98} using RXTE/PCA, ASCA and Ginga data), 
and in all cases X-ray determined masses where 
substantially higher than masses derived from optical  timing or spectroscopy.
The classical example is XY Ari, an eclipsing IP (which is considered to
be the most suitable object because the uncertainty in the inclination is
very small): the RXTE spectrum yields 1.22 \msun, the ASCA data 
1.27-1.40 \msun\ depending on GIS, SIS or a combination of the two detectors,
and the Ginga data result in 1.3 \msun, while optical methods yield 
0.78--1.03 \msun\ \cite[][]{ramsay98}.

Observations above 20\,keV are crucial as they provide a clean signal
not contaminated by other system components such as the 
accretion disc or the  main-sequence star and it avoids complications
due to cold or warm absorbers as well as fluorescence (line) emission.
Moreover, at or above this energy, effects of 
absorption in the post-shock region are negligible.
The only effect to take into account is reflection which is known
to be present in magnetic cataclysmic variables due to the
detection of the 6.4 keV emission line.
The only disadvantage was the weakness of the X-ray flux. Thus, a
sensitive instrument must be used for this kind of observation.
The {\em Swift} Burst Alert Telescope \cite[BAT,][]{barthelmy04}, with an
unprecedented sensitivity in the 15--200\,keV range, opened
a new window for the study of the hard X-ray sky, combining high sensitivity and
all-sky coverage.

The present paper is organized as follows: in section 2 we introduce the 
observations. The results of the measurement 
can be found in section 3 which is divided in three subsections to discuss 
first the spectral analysis, second the 
accretion-rate dependence observed in GK Per, 
and finally the comparison with previous methods of mass determination
(both with X-ray and non X-ray methods). 
The conclusions are given in section 4.

\section{Data and models}

\subsection{The {\em Swift}/BAT X-ray survey}

BAT is a coded mask telescope with a wide field of view aperture (FOV, 
120$^\circ$ $\times$ 90$^\circ$ partially
coded)  sensitive in the energy range of 15--195\,keV. 
BAT continuously monitors  the hard X-ray sky surveying gamma-ray bursts. 
Results of the BAT survey \citep{markwardt05,marco7a} 
show that BAT reaches a sensitivity of $\approx$1 mCrab in a
1 Ms exposure. With its sensitivity and the large amount of exposure 
already accumulated over the
whole sky, BAT is a very good instrument for studying IPs above 20 keV.

In the analysis presented here we use BAT data to derive spectra
of known IPs. This was done by cross-correlating our list of source
candidates with catalogs of IPs.

For the source detection step, we used all the available BAT data taken 
from January 2005 to March 2007.
 An energy range of 15-55\,keV was chosen. 
The lower limit is given by the energy
threshold of the detectors. The upper limit is taken to avoid the presence 
of strong background lines which could worsen the overall sensitivity. 
The data screening was performed
analogously to the method presented in \cite{marco7a}. The final all-sky image
is obtained as the weighted average of all the shorter observations. 
The average exposure
time in our image is 3 Ms, with 1.3 Ms  minimum and 5 Ms   
maximum exposure times. The final image shows  Gaussian normal noise and 
we identified source
candidates as excesses above the 3$\sigma$ level. All the candidates are 
then fit with the BAT point
spread function (using the standard BAT tool batcelldetect) to derive the 
best source position. The BAT position error is between 3-7 arcmin,
depending on the source intensity (see \cite{marco7a}).
The identification of IPs from this detection list was done using
the online catalog maintained by 
K. Mukai\footnote{http://asd.gsfc.nasa.gov/Koji.Mukai/iphome/iphome.html,
version of March 2007} and allowing for each source's position error.

Out of the 29 known IPs to date,
{\it Swift}/BAT detects 22 objects  (see Table \ref{tbl:werte2}).
The sources that fall below the BAT sensitivity
(AE Aqr, HT Cam, DW Cnc, UU Col, DQ Her 
and WX Pyx) are likely either very far away, or are in a very low
accretion state.
All six sources except HT Cam (RXTE detection; \cite{revnivtsev04})
were also not detected in RXTE or Integral/IBIS observations. 

For each of the detected IPs, we extracted a 15--195\,keV spectrum
with the method described in \cite{marco7b}.
Here we recall the main steps.
For a given source, we extract a spectrum from each observation for 
which the source is in the field of view. These spectra are corrected for 
residual background contamination and for vignetting. The per-pointing 
spectra are then (weighted) averaged to produce the final source spectrum. 
Thus, the final spectrum represents the average source emission over the 
entire survey period (2.5 years) although individual coverage intervals 
may differ.

We also extracted the light curves of all detected IPs. 
In most of the cases, the sources
are too faint to assess any variability. The only exception is 
GK Per (see Fig.~\ref{fig:gkperlcs}) and this source will be discussed
in detail in Section \ref{sec:ass}.

%%%%%%%%%%%%%%%%%%%%%%%%%%%%%%%%%%%%%%%%%%%%%%%%%%%%%%%%%%%%%%%%%%%%%%%%%%%%%%

\begin{figure}
\centering
 \includegraphics[width=0.43\textwidth,angle=270,scale=0.8]{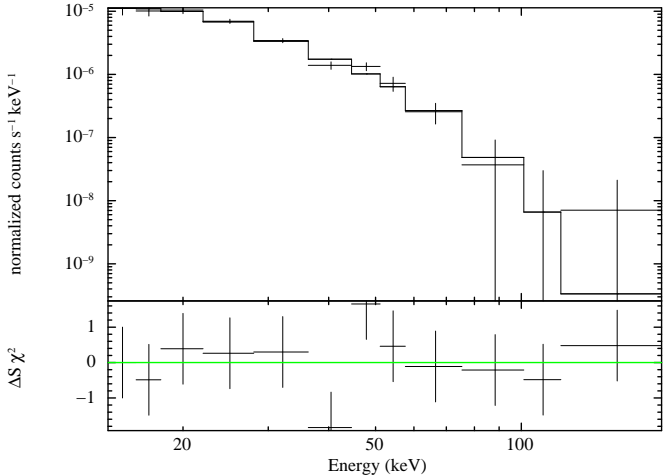}
 \caption{{\em Swift}/BAT spectrum of GK Per, averaged over 2.5 yrs, fitted 
  with the model described in \cite{suleimanov05}.}
 \label{fig:gkpersp}
\end{figure}

\subsection{Spectral model}

The main cooling mechanism in the post-shock region is through 
bremsstrahlung, thus the simplest method is to fit a Bremsstrahlung
spectrum to the data. The structure of the post-shock region was 
investigated first by \cite{aizu73}, who provided a conversion formula 
to determine the white dwarf mass based on the effective
temperature of the  Bremsstrahlung spectrum. However,
in order to estimate the mass of the  accreting white dwarf more 
accurately, one needs a more careful modeling of the 
temperature and emissivity distribution in the post-shock region.
\cite{WoelkBeuer96} took into account cyclotron cooling, which can be 
important for polars, but was shown to be unimportant for objects
with surface magnetic fields less than 10 MG (like IPs). 
\cite{Frank02} derived a simple analytical model under the assumption
of constant pressure in the post-shock region. 
\cite{suleimanov05} extended this model to take into account how the
pressure grows towards the white dwarf surface. The emergent spectrum
thus is the sum of many local Bremsstrahlung spectra with their appropriate
temperatures and pressures. While this model does not include
reflection, \cite{cropper98} have shown that including the reflection
changes the fit parameters and the corresponding white dwarf masses
only slightly.
The main difference between the model of \cite{suleimanov05}
and that of Aizu (1973) is that 
\cite{suleimanov05} uses the cooling functions of 
\cite{sutherland93}, which takes thermal Bremsstrahlung, lines and b-f 
transitions into account.

In the following we have used the model of \cite{suleimanov05}.
The free parameters are the mass of the white dwarf and
a normalization constant. Since we fit spectra only above 15 keV,
we do not apply any foreground absorption.

%%%%%%%%%%%%%%%%%%%%%%%%%%%%%%%%%%%%%%%%%%%%%%%%%%%%%%%%%%%%%%%%%%%%%%%%%

\section{Results}

\begin{figure*}[htb]
   \includegraphics[width=0.34\textwidth]{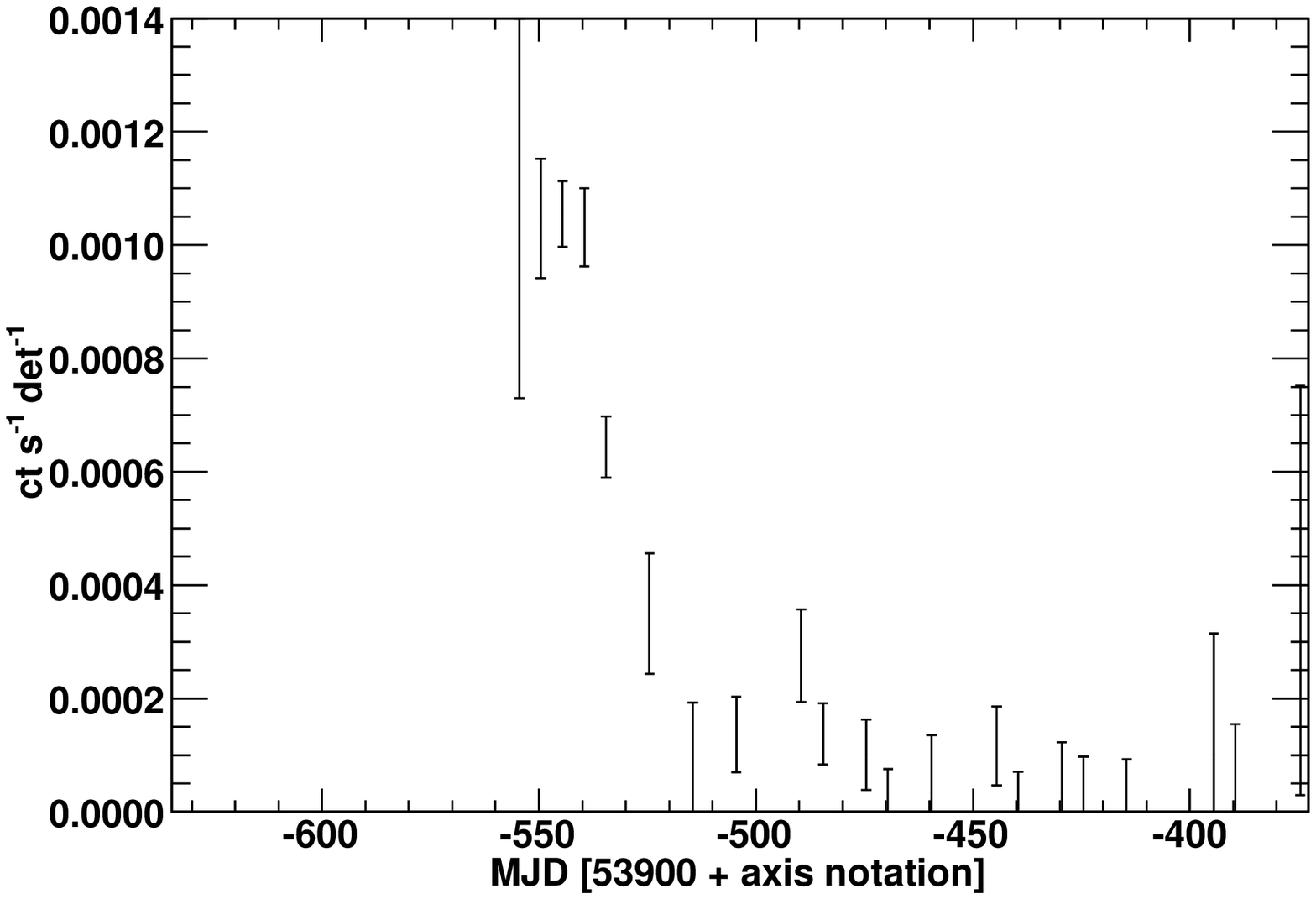}
   \includegraphics[width=0.34\textwidth]{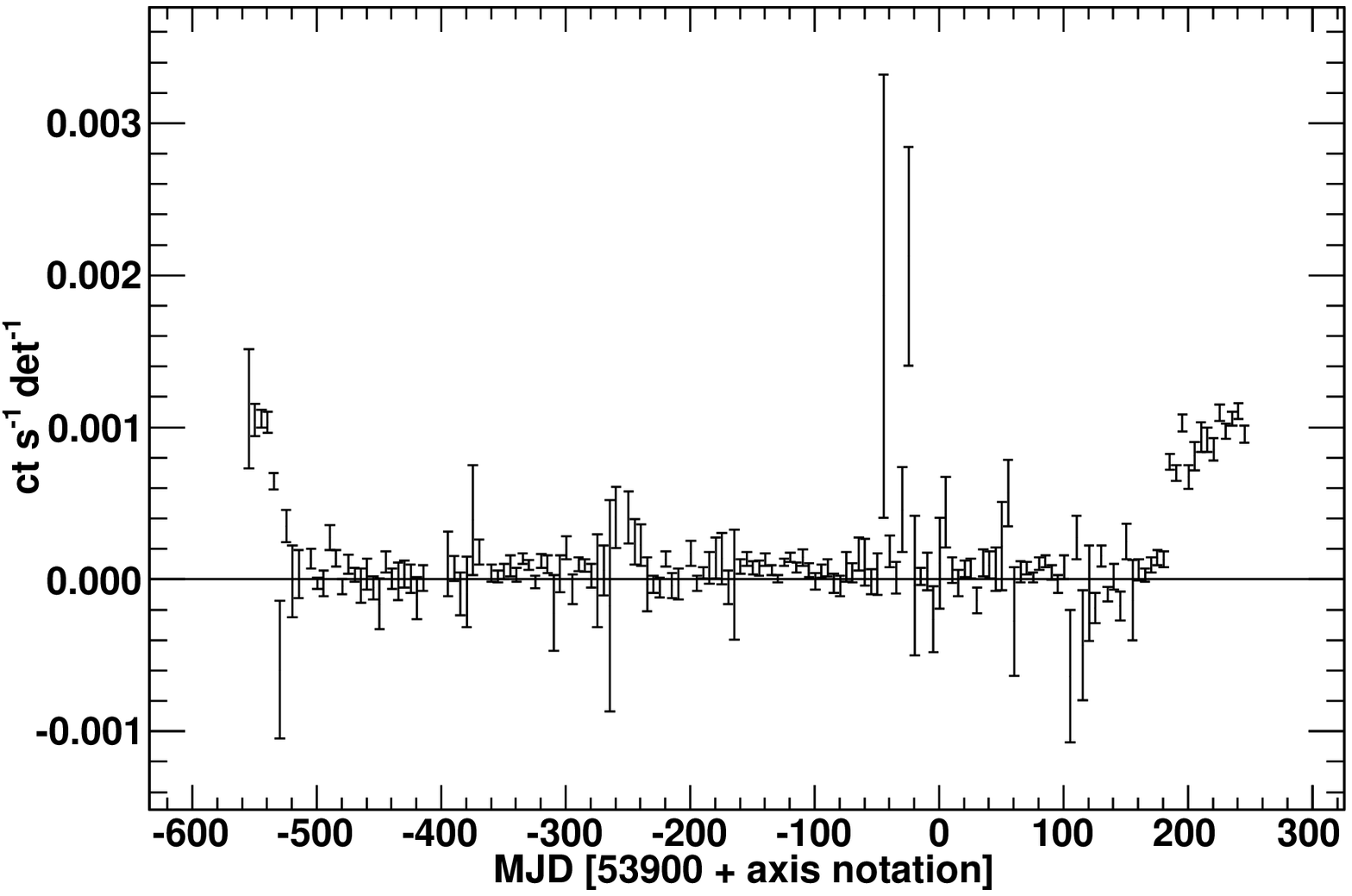}
   \includegraphics[width=0.34\textwidth]{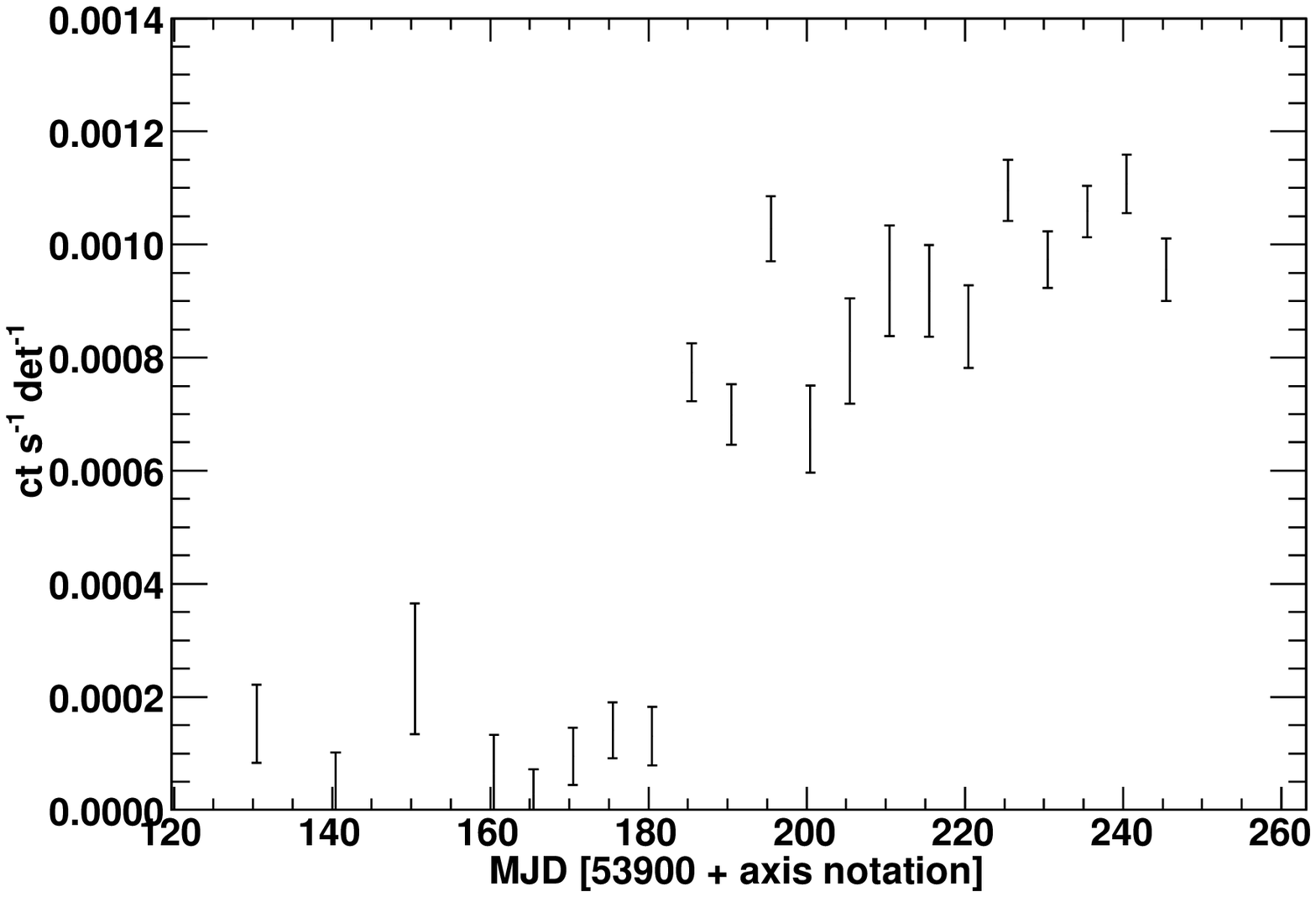}
  \caption{Light curves of GK Per in the 15--195\,keV band with a 
  binning of 5 days. The middle panel shows the full observation 
  while the left panel
  shows the outburst at the beginning and the right panel that 
  at the end of the covered observing period.}
  \label{fig:gkperlcs}
  \bigskip
\end{figure*}

%%%%%%%%%%%%%%%%%%%%%%%%%%%%%%%%%%%%%%%%%%%%%%%%%%%%%%%%%%%%%%%%%%%%%%%%%%%%%%
\subsection{Spectral analysis}
\label{sec:spec}

In order to derive a WD mass estimate we fitted  the model of 
\cite{suleimanov05} to the BAT IPs  using XSPEC.
An example is given in Figure \ref{fig:gkpersp}.
The detection of the IP emission up to $\sim$100\,keV allows an accurate 
estimate of the bremsstrahlung temperature.
Since the X-ray spectrum depends primarily on the WD mass,
using the model of \cite{suleimanov05} we are able to determine
directly an estimate (and its uncertainty) of the WD mass.

With the exception of MU Cam, all spectral fits show 
acceptable reduced $\chi^2$ (see Table \ref{tbl:werte} which include
the errors at the 90\% confidence level; the number of degrees of freedom 
is always 10, based on 12 channels and 2 fit parameters).
We refrained from attempting to fit more sophisticated models.
Note that, in all the cases, a power law does not produce a good fit
because of the clear curvature of the data.

\begin{table*}
 \caption{Detected IPs with masses from XSPEC fit with the model from 
\cite{suleimanov05}. 
  The temperature is from a bremsstrahlung fit, and is provided for 
  comparison.}
\begin{tabular}{lrlrclcr}
 \hline 
  \noalign{\smallskip}
   Name & kT~~~~ & ~~~Mass &  F$_{14-195}$~~~ &  Distance$^A$  & L$_{14-195}$  & \mdot & $\chi^2_{\rm red}$ \\
        & (keV)~~ & ~~~(\msun)  & (10$^{-11}$erg/cm$^2$/s) & ~~(pc) & (10$^{32}$ erg/s) & (10$^{16}$ g/s) & \\
  \noalign{\smallskip}
   \hline 
  \noalign{\smallskip}
 FO Aqr  & 15.2$\pm$1.6  & 0.61 $\pm$ 0.05 & 4.7 $\pm$ 0.5 & 400$^1$ & 9.00 &29.0 &1.4  \\
 XY Ari  & 29.1$\pm$6.2  & 0.96 $\pm$ 0.12 & 2.3 $\pm$ 0.4 & 270$^2$    &  2.01 &2.8 &1.3  \\
 V405 Aur& 22.1$\pm$4.2  & 0.89 $\pm$ 0.13$^C$ & 2.2 $\pm$ 0.5 &  - & - & &0.5  \\
 MU Cam  & 19.6$\pm$4.7  & 0.74 $\pm$ 0.13 & 1.3 $\pm$ 0.3 &  - & - & &1.7  \\
 V709 Cas& 30.0$\pm$2.5  & 0.96 $\pm$ 0.05$^C$ & 7.2 $\pm$ 0.6 &  - & - & &1.0  \\
 V1025 Cen&10.8$\pm$7.7  & 0.46 $\pm$ 0.31 & 0.4 $\pm$ 0.3 &  - & - & &0.6  \\
 BG CMi   &18.9$\pm$5.5  & 0.67 $\pm$ 0.19 & 2.0 $\pm$ 0.4 &  700$^3$    & 11.73 &32.3 &0.7  \\
 TV Col   &21.6$\pm$2.4  & 0.78 $\pm$ 0.06 & 5.0 $\pm$ 0.5 &  386$^4$    &  8.92 & 18.8&0.9  \\
 TX Col   &17.0$\pm$3.2  & 0.67 $\pm$ 0.10 & 1.7 $\pm$ 0.3 &  550$^5$    &  6.15 & 16.9&0.7  \\
 V2306 Cyg&20.5$\pm$5.7  & 0.77 $\pm$ 0.16 & 1.2 $\pm$ 0.3 &  - & - & &1.0  \\
 DO Dra   &12.0$\pm$2.6  & 0.50 $\pm$ 0.11$^C$ & 1.0 $\pm$ 0.3 &  155$^6$    &  0.29 &1.3 &0.7  \\
 PQ Gem   &16.8$\pm$2.8  & 0.65 $\pm$ 0.09 & 2.5 $\pm$ 0.4 &  - & - & &1.0  \\
 EX Hya   &19.4$\pm$4.4  & 0.66 $\pm$ 0.17 & 2.1 $\pm$ 0.5 &  64.5$^7$   &  0.10 & 0.28 & 0.4  \\
 NY Lup   &39.1$\pm$4.3  & 1.09 $\pm$ 0.07 & 7.1 $\pm$ 0.6 & 690$^8$    & 40.45 & 25.3 &1.1  \\
 V2400 Oph&21.1$\pm$2.8  & 0.81 $\pm$ 0.10 & 4.3 $\pm$ 0.6 &  - & - & &1.0  \\
 GK Per   &26.2$\pm$5.4  & 0.90 $\pm$ 0.12$^{B,C}$$\!\!$& 2.8 $\pm$ 0.5 &  340$^9$    & 10.24 & 16.2 &0.8  \\
 AO Psc   &13.1$\pm$1.9  & 0.55 $\pm$ 0.06 & 2.8 $\pm$ 0.4 & 420$^{10}$   & 10.13 &38.6 & 0.7  \\
 V1223 Sgr&17.0$\pm$1.3  & 0.65 $\pm$ 0.04 & 8.5 $\pm$ 0.8 & 527$^{11}$ & 28.25 & 82.0 &1.0  \\
 V1062 Tau&18.3$\pm$5.8  & 0.72 $\pm$ 0.17 & 1.6 $\pm$ 0.4 &  500$^{12}$ &  4.79 & 11.6 & 0.4 \\
 RX J2133 &27.1$\pm$2.9  & 0.91 $\pm$ 0.06 & 4.6 $\pm$ 0.4 &  $>$600$^{13}$   & - & &0.6  \\
 Swift J0713&21.7$\pm$4.1 & 0.79 $\pm$ 0.11 & 2.5 $\pm$ 0.4 &  - & - & &0.9  \\
 IGR J00234&22.8$^{+30.0}_{-11.5}$ & 0.85 $\pm$ 0.39 & 0.7 $\pm$ 0.3 &  - & - & &0.8  \\
 IGR J17303&37.1$\pm$4.4 & 1.08 $\pm$ 0.07 & 6.6 $\pm$ 0.6 & - & - & &1.1  \\
  \noalign{\smallskip}
  \hline 
  \noalign{\smallskip}
 \end{tabular}

 $^A$ References to the distances: 1. \cite{mchardy87}; 2. \cite{littlefair01};
   3. \cite{berriman87}; 4. \cite{mcarthur01}; 5. \cite{buckley89}; 
   6. \cite{mateo91}; 7. \cite{beuermann03}; 8. \cite{demartino06}; 
   9. \cite{warner87}; 10. \cite{hellier91}; 11. \cite{beuermann04}; 
  12. \cite{szkody96}; 13 \cite{Bonnet-Bidaud06}. \\
 $^B$ All numbers correspond to the measured spectrum during the low-flux
     period; but see section 3.2. \\
 $^C$ The derived mass is likely a lower limit due to the fast rotation
     of the WD; see section 3.1.
 \label{tbl:werte}
 \bigskip
\end{table*}

%%%%%%%%%%%%%%%%%%%%%%%%%%%%%%%%%%%%%%%%%%%%%%%%%%%%%%%%%%%%%%%%%%%%%%%%%%%%%%%
%%%%%%%%%%%%%%%%%%%%%%%%%%%%%%%%%%%%%%%%%%%%%%%%%%%%%%%%%%%%%%%%%%%%%%%%%%%%%%%

\subsection{Accretion-rate dependence of GK Per}
\label{sec:ass}

In the simplistic case, it is assumed that the accreted mass falls 
freely from infinity
onto the white dwarf. In reality, however, the matter does not come from
infinity, but falls from the Alfv\'en radius 
R$_A$, which in turn depends on the mass accretion rate 
(from \cite{suleimanov05}):

\begin{equation}
 \frac{R_A}{R_{\rm WD}} \approx 
2.3\left(\frac{\dot{M}}{10^{20} {\rm g/s}}\right)^{-2/7}\left(\frac{M_{\rm WD}}{M_{\odot}}\right)^{-1/7}\left(\frac{R_{\rm WD}}{10^9 {\rm cm}}\right)^{5/7}\left(\frac{B}{10^6 {\rm G}}\right)^{4/7} 
 \label{eq:alfven}
\end{equation}

Thus, including this dependency, the white dwarf mass is:
\begin{equation}
M_{\rm WD}(R_A)=M_{\rm WD}(\infty)\left(1-\frac{R_{\rm WD}}{R_{\rm A}}\right) .
\label{eq:mwirk}
\end{equation}

Therefore, the higher the accretion rate, the smaller is R$_{\rm A}$, and
thus the smaller the mass of the white dwarf which is deduced
under the assumption of infinite origin. 

There is yet another dependency which makes a statement on the
deduced mass relative to the real mass somewhat uncertain.
Accretion is only possible if $R_A<R_{\Omega}$ \citep{warner95}, 
where $R_{\Omega}$ is 
the "corotation" radius, i.e. the radius where the Keplerian angular 
velocity is equal to the rotation velocity of the WD
(otherwise the mass would be thrown out of the system):

\begin{equation}
\frac{R_{\Omega}}{R_{\rm WD}}=2.3 P_{p}^{2/3}{\rm (min)}\left(\frac{M_{\rm WD}}{M_{\odot}}\right)^{1/3}\left(\frac{R_{\rm WD}}{10^9 {\rm cm}}\right)^{-1} .
\label{eq:corot}
\end{equation}

Thus, fast rotating systems require higher accretion rates at the same
magnetic field strength, and we measure 
smaller masses than they have in reality. Because of this,
we think we measure only a lower limit for the mass of the following four 
systems (with their WD rotation period in parenthesis): 
GK Per  (5.85 min), V709 Cas  (5.12 min), 
DO Dra  (8.83 min), V405 Aur  (9.09 min).

Fig. \ref{fig:gkperlcs} reveals that GK Per had an outburst
 at  the beginning (January 2005) as well as at the end 
(Dec. 2006 -- $>$Mar 2007) 
of the covered  observing period. GK Per is well known for its 
periodic outbursts, which take place every three or four years 
\citep{sabbadin83, simon2002}.  
\cite{brat06} mention an increased optical brightness in Dec. 2006, 
about 0.6 mag above its usual quiescent level, which is consistent
with our detection of increased X-ray emission.
Since GK Per is among the brightest IPs in our sample, we are able
to split the data into three different periods, namely the low-flux
period as well as the two outbursts, and derive
the mass separately for these periods.

\begin{table}
\centering
 \caption{Comparison of mass estimates for different sections of the 
  light curve of GK Per. 
 }
 \begin{tabular}{llllr}
 \hline 
 \noalign{\smallskip}
 \textbf{Part of light curve}  & \textbf{Mass (M$_{\odot}$) } & 
     \textbf{$\chi^2_{\rm red}$}  &  \\
  \noalign{\smallskip}           
  \hline 
  \noalign{\smallskip}  
 Low-flux period, without outbursts & 0.90$\pm$0.12 & 0.78&  \\
 Jan. 2005 outburst                 & 0.74$\pm$0.05 & 1.66&  \\
 Mar. 2007 outburst                 & 0.67$\pm$0.03 & 1.24 &  \\
  \noalign{\smallskip}  
 Sum of the total period            & 0.78$\pm$0.05 & 0.79  &  \\
  \noalign{\smallskip}  
 \hline 
 \end{tabular}
  \label{tbl:highlow}
\end{table}

The results in Table \ref{tbl:highlow} show that we 
measure lower masses at higher flux. While our data are not sufficient
to demonstrate that the mass estimates quantitatively follow that
expected on the basis of the expected change in R$_A$, the change
is suggestive of being due the
dependence on the accretion rate according to equations \ref{eq:alfven} and 
\ref{eq:mwirk}.
Note the systematically higher mass estimate if we were not been able
to distinguish the different flux levels.
Only the mass derived during the low-flux (low-accretion)
period of GK Per should be considered -- the last row in Table \ref{tbl:highlow}
is for illustrative purposes only.
The values as derived during the outbursts 
are similar to those of \cite{suleimanov05}, suggesting that
those data were also taken during periods of increased accretion.
We finally note that for all other sources of Table \ref{tbl:werte}
the mass estimates are unlikely to be affected by high accretion rates,
i.e. the source being in outburst.

\begin{figure}[htb]
%\centering
  \includegraphics[width=0.52\textwidth]{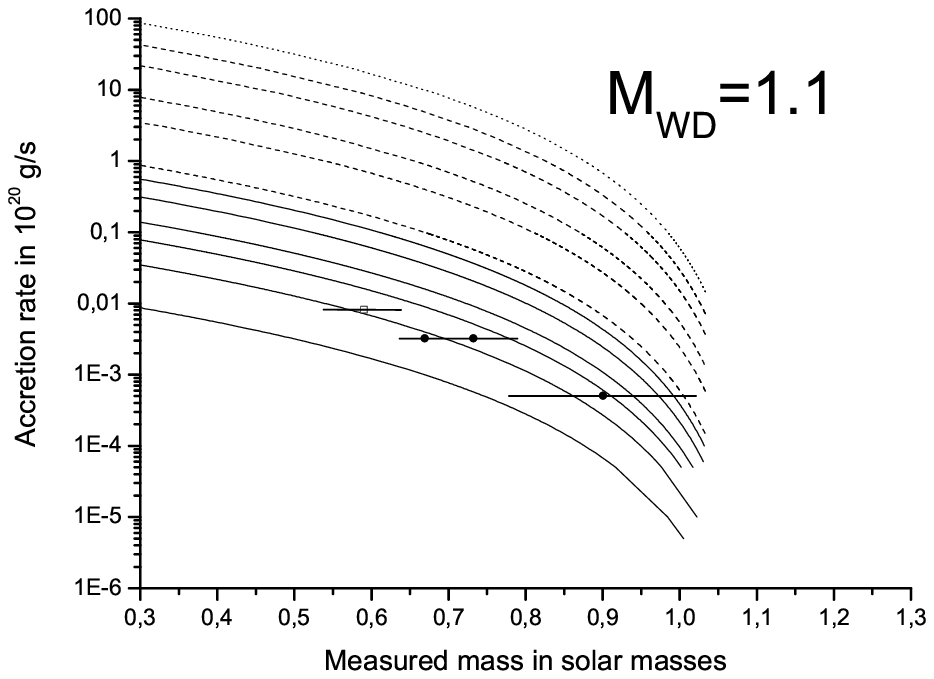}
  \includegraphics[width=0.52\textwidth]{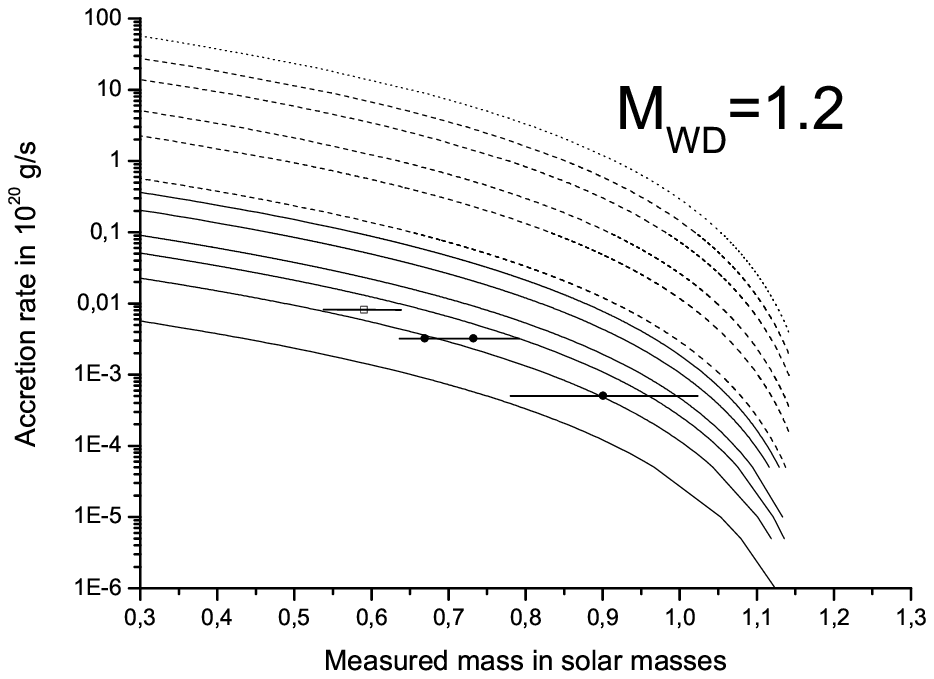}
 \caption{Relation between measured mass and accretion rate for different 
   magnetic fields for M$_{WD}$=1.1 M$_{\odot}$ and M$_{WD}$=1.2 M$_{\odot}$. 
   The values of this work are shown (filled circles)
   with accretion rates from \cite{warner95}.
   The parallel curves are for different values of the magnetic field,
   namely 0.1, 0.2, 0.3, 0.4, 0.6, 0.8, 1, 2, 3, 5, 7 and 10 MG, respectively
   (from bottom to top).
   The open square represents the measured mass and its corresponding 
   accretion rate derived by \cite{suleimanov05}. The accretion rate 
   has its first limit from the constraint that the system must have an 
   accretion rate lower 
   than 0.01$\cdot10^{20} g/s$ at quiescence, because of its life time. 
	}
   \label{fig:mmm}
\end{figure}

Assuming that the magnetic field (and the mass) does not change
significantly,
we attempted to determine the accretion rate via equations \ref{eq:alfven},
\ref{eq:mwirk} and \ref{eq:corot}, using the rotation period of 
351 sec  \cite{watson85}.
In Fig. \ref{fig:mmm} the three mass estimates from Table \ref{tbl:highlow} are 
shown together with the generally adopted values for the mass accretion rate
of GK Per in its different states 
\citep{warner95}, namely 4$\times 10^{16}$ g/s at quiescence and 
23$\times 10^{16}$ g/s in outburst. We also add
the measurement of M = 0.6 \msun\ from \cite{suleimanov05}. 
In order to be consistent, all four  measurements should
fall onto one particular line for the magnetic field of the white dwarf.
Testing white dwarf masses between 0.6--1.3 M$_{\odot}$, we find that
the mass of the  white dwarf in GK Per must be definitely higher than 
0.9 M$_{\odot}$. A mass of 1.15 M$_{\odot}$ provides the best match, and
we present the  graphs for 1.1 and 1.2 M$_{\odot}$ to show how sensitive 
the mass estimate is (note, however, that this hinges on the
assumption of the accretion rate).
For 1.1 M$_{\odot}$ the data points are flatter than
the corresponding curve of B = 0.2 MG, while for 1.2 M$_{\odot}$ the data 
points are already steeper than the nearest B=const. curve.
The online material provides the remaining figures with white dwarf masses of 
0.9 M$_{\odot}$, 1.0 M$_{\odot}$ and 1.3 M$_{\odot}$. 
Our accretion rate corrected mass of the white dwarf of GK Per of
1.15 M$_{\odot}$ compares well with the lower limit of 0.9 M$_{\odot}$
derived by Morales-Rueda et al. (2002).
However, the implied magnetic field of 0.2 MG is at the low end of 
what is believed
to be the canonical magnetic field strength of IPs, namely 1--10 MG
\citep{deMartino2004}.

The canonical way of deriving the accretion rate is via the
X-ray luminosity, i.e. using the relation 
$L_{\rm X} = 0.1 \times \epsilon \times G \times M_{\rm WD} 
              \times \dot M / R_{\rm WD}$, 
where $\epsilon$
is the bolometric correction to the X-ray luminosity in the 14-195 keV band,
the factor 0.1 accounts for the ratio of X-ray to total accretion luminosity,
and we assume
the conversion efficiency of gravitational binding energy to radiation 
being equal to 1 for WD surface accretion, as well as isotropic emission.
For a bremsstrahlung spectrum of 20 (40) keV, $\epsilon$ = 0.33 (0.53).
Using our mass estimate of 0.90 (1.15) \msun\ and the corresponding radius
of 6200 (3500) km (according to the relativistic Fermi gas), we obtain
\mdot = 1.7 (0.7) $\times 10^{16}$ g/s.
This is somewhat lower than according to the typical values
used above, and, if true, would suggest an even lower magnetic field.
It seems obvious that the combination of the accretion rate estimate from
$L_{\rm X}$ with the dependencies according to eqs. 1--3 is not fully
consistent with our general belief of B $\sim$ 1-10 MG in IPs.

\begin{figure}[htb]
 \centering
 \includegraphics[width=0.53\textwidth]{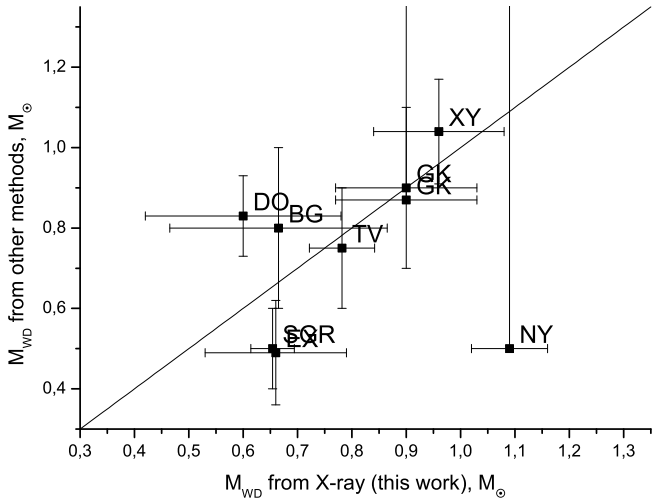}
 \caption{Comparison of the measured white dwarf masses with the ones 
  from other methods  (values from Table \ref{tbl:vgl}). 
  The source names are abbreviated; the labels of V1223 Sgr and EX Hya
  overlap.}
 \label{fig:vggl}
\end{figure}

\subsection{Comparison with other methods of mass determination} 
\label{sec:vergleich}
 
The canonical method of determining the WD mass in CVs is by measuring the
radial velocities of optical emission lines. However, this method
suffers from two main problems. First, the uncertainty in the inclination of
the system leads to a wide range in mass. Second, in magnetic systems
the velocities along the collimating field lines dominate the 
radial velocities.

An alternative method is to measure the mass ratio and infer the mass
of the secondary by modeling the infrared light curves 
\citep[e.g.][]{Allan96}.
Again, the uncertainty in the inclination limits the accuracy of the mass 
determination.

Eclipsing systems offer a unique opportunity to measure the mass
more accurately due to the relatively small uncertainty in the inclination
of the binary system. Moreover, in eclipsing systems the radius of the white
dwarf can be determined, thus providing yet another method (though rarely
applicable due to the few eclipsing systems) to determine the WD mass.

\subsubsection{The whole sample} 
 
%\begin{sidewaystable} 
\begin{table*}
\centering
  \caption{Comparison of our mass estimates with those from previous 
    X-ray observations}
 \begin{tabular}{llllllllr}
% \noalign{\smallskip}
 \hline 
 \noalign{\smallskip}
   \textbf{Name}  & \textbf{BAT$^1$} & \textbf{RXTE$^2$} &  \textbf{RXTE$^3$} & \textbf{GINGA$^4$} & \textbf{ASCA$^5$} & \textbf{Other methods$^6$} & \textbf{References$^7$} \\
                   & M$_{\odot }$ & M$_{\odot }$ & M$_{\odot }$ & M$_{\odot }$ & M$_{\odot}$ & M$_{\odot}$ &     \\
 \noalign{\smallskip}
 \hline 
 \noalign{\smallskip}
 FO Aqr           & 0.61$\pm$0.05 & 0.60$\pm$0.05 & 0.88 & 0.92 & 1.05 &               &            \\
 XY Ari                 & 0.96$\pm$0.12 &               & 0.97 &      &      & 1.04$\pm$0.13 & [1] \\
 V405 Aur         & 0.89$\pm$0.13 & 0.90$\pm$0.10 & 1.10 &      & $>$0.54&               &            \\
% MU Cam                                   & 0.74$\pm$0.13 &               &      &      &      &               &            \\
 V709 Cas         & 0.96$\pm$0.05 & 0.90$\pm$0.10 & 1.08 &      &      &               &            \\
 V1025 Cen            & 0.46$\pm$0.31 &               & 0.60 &      &      &               &            \\  
 BG CMi                                   & 0.67$\pm$0.19 & 0.85$\pm$0.12 & 1.20 & 1.09 &      & 0.8$\pm$0.2   & [2] \\    
 TV Col                 & 0.78$\pm$0.06 & 0.84$\pm$0.06 & 0.96 & 1.30 & 0.51 & 0.75$\pm$0.15 & [3] \\
 TX Col           & 0.67$\pm$0.10 & 0.70$\pm$0.30 & 0.73 & 0.48 & 0.66 &               &            \\
% V2306 Cyg        & 0.77$\pm$0.16 &               &      &      &      &               &            \\  
 DO Dra           & 0.50$\pm$0.11 & 0.75$\pm$0.05 &      &      &      & 0.83$\pm$0.10 & [4] \\
 PQ Gem           & 0.65$\pm$0.09 & 0.65$\pm$0.20 &      & 1.29 &      &               &            \\
 EX Hya           & 0.66$\pm$0.13 & 0.50$\pm$0.05 & 0.45 & 0.46 & 0.48 & 0.49$\pm$0.13 & [5] \\
 NY Lup                 & 1.09$\pm$0.07 &               &      &      &      & $>$0.5          & [6] \\
 V2400 Oph        & 0.81$\pm$0.10 & 0.59$\pm$0.05 & 0.71 &      & 0.68 &               &            \\
 GK Per                 & 0.90$\pm$0.12 & 0.59$\pm$0.05 &      &      & 0.52 & 0.9$\pm$0.2, $>$0.87 & [7], [8] \\
 AO Psc           & 0.55$\pm$0.06 & 0.65$\pm$0.05 & 0.60 & 0.56 & 0.40 &               &            \\
 V1223 Sgr            & 0.65$\pm$0.04 & 0.95$\pm$0.05 & 1.10 &      & 1.28 & 0.4-0.6         & [2]   \\
 V1062 Tau            & 0.72$\pm$0.17 & 1.00$\pm$0.20 & 0.90 &      &      &               &            \\
\noalign{\smallskip}
 \hline 
 \noalign{\smallskip}
  \end{tabular}  

References: 1: this work, BAT; 2: \cite{suleimanov05};
  3: \cite{ramsay00}; 4: \cite{cropper99}; 5: Ezucka,Ishida (1999); 
  6: Other methods of determination of mass and the last column 
     are the references of them.
  7: References: [1] \cite{hellier97},
                 [2] \cite{penning85},
                 [3] \cite{hellier93},
                 [4] \cite{haswell97},
                 [5] \cite{hoogerwerf04},
                 [6] \cite{demartino06},
                 [7] \cite{crampton86},
                 [8] \cite{morales02}.
 \label{tbl:vgl}
\end{table*}
%\end{sidewaystable}

In Table \ref{tbl:vgl} and Fig. \ref{fig:vggl}
we compare the mass estimates obtained by our
analysis to previous mass estimates via radial velocity studies. 
In general, there is good agreement, with no systematic effect visible.
A detailed comparison with respect to individual sources 
is  given below.

While we compare our results to those obtained with IBIS onboard INTEGRAL
(see below), we choose not to discuss the differences (Table \ref{tbl:vgl})
of our results to those obtained by Ginga, ASCA and RXTE
due to the systematic, and unexplained, differences for XY Ari 
\citep{ramsay98}. In the next section we discuss each object.

\subsubsection{Discussion of individual objects}

The mass of 0.96$\pm$0.12 M$_{\odot}$ for \textbf{XY Ari} matches very well
the value of 1.04$\pm$0.13 M$_{\odot}$ from \cite{hellier97}. 
This is particularly encouraging given the systematic differences
in previous estimates as discussed in \cite{ramsay98} -- see the Introduction.
Since XY Ari is the only eclipsing IP, its optically determined mass
is generally assumed to be secure. The fact that our mass estimate
matches so well suggests that our analysis does not suffer from
major systematic uncertainties.

\cite{penning85} derives a mass range of 0.6--1.0  M$_{\odot}$
for the WD in \textbf{BG CMi}. Our value of 0.72$\pm$0.14 M$_{\odot}$ 
falls  in that range.

For the system \textbf{TV Col}, \cite{hellier93} determined a mass of 
0.75$\pm$0.15 M$_{\odot}$, which agrees very well with our estimate of
0.78$\pm$0.06 M$_{\odot}$.

In \textbf{DO Dra} our WD mass estimate is 0.50$\pm$0.11 M$_{\odot}$. 
\cite{haswell97} find a mass of 0.83$\pm$0.10 M$_{\odot}$, which is 
substantially different to our value.  
\cite{mateo91} also find a mass of 0.83$\pm$0.10 M$_{\odot}$. 
The difference could be the result of the short spin period of the WD 
(see subsection \ref{sec:ass}), thus making our estimate a lower limit.
However, our mass estimates of the other WDs with short spin period
are not as far below the optical estimates.

Using data of the first months of BAT observations, a significantly higher
mass (0.83$\pm$0.17 \msun) is deduced for \textbf{EX Hya} than for 
the rest  of the time - although the count rate does not differ
in these periods. 
If  the first 100 days are ignored,  a mass of 0.66$\pm$0.13 M$_{\odot}$ 
is derived. While we cannot completely exclude a statistical fluctuation
which leads to a change in the spectrum, 
one possible physical reason for the strange behavior could be that 
this system has a much lower mass accretion rate than the others. 
In the model of \cite{suleimanov05} cyclotron cooling plays a role 
for a $<$ 0.01 g/s/cm$^2$, so if the accretion rate were lower,
this could provide an explanation.
Hoogerwerf et al. (2004) measured a mass for 
the WD in EX Hya of 0.49$\pm$0.13 M$_{\odot}$, which is comparable within 
the errors with our value.

In the system \textbf{NY Lup} we obtain a mass of 1.12$\pm$0.06 M$_{\odot}$. 
This value is consistent with the lower limit of 0.5 M$_{\odot}$, which 
Martino et al. (2006) determined. 

For \textbf{V1223 Sgr} a WD mass of 0.65$\pm$0.04 M$_{\odot}$ was 
derived.
\cite{penning85} gives a range of 0.4-0.6 M$_{\odot}$, which fits our value.

The accretion-corrected mass of 1.15$\pm$0.15 M$_{\odot}$ for \textbf{GK Per} 
is within the error bars of the values from other methods: 
0.9$\pm$0.2 M$_{\odot}$ from Crampton et al. (1986) and the lower 
limit M$>$0.87$\pm$0.24 M$_{\odot}$ from 
\cite{morales02}.

%%%%%%%%%%%%%%%%%%%%%%%%%%%%%%%%%%%%%%%%%%%%%%%%%%%%%%%%%%%%%%%%%%%%%%%%%%%%%%%

\begin{table*}
\centering
 \caption{Comparison of masses and effective temperatures with a 
  remsstrahlung model from BAT (this work, model: \cite{suleimanov05}) and 
  INTEGRAL/IBIS by \citep{barlow06} and \citep{landi08}. 
  1: Masses, derived from the temperatures with the model of \cite{aizu73}.
  2: Bremsstrahlung temperature. }
 \begin{tabular}{llllllll}
 \hline 
 \noalign{\smallskip}
  Name  & \multicolumn{2}{c}{BAT/this work}  & \multicolumn{2}{c}{IBIS/Barlow}   & \multicolumn{2}{c}{IBIS/Landi}  \\
        & mass (M$_{\odot}$) & kT  (keV) & mass (M$_{\odot}$)$^1$ & kT  (keV)$^2$  & mass (M$_{\odot}$)$^1$ &  kT  (keV)$^2$\\
 \noalign{\smallskip}
  \hline 
 \noalign{\smallskip}
 FO Aqr     & 0.61$\pm$0.05 & 15.2$^{+1.7}_{-1.5}$  &       --     &  --        &
   1.06$^{+0.38}_{-0.51}$ & 29.7$^{+70.1}_{-16.6}$ \\
 MU Cam     & 0.74$\pm$0.13 & 19.6$\pm$4.7  & 0.37$\pm$0.17 &8.1$\pm$4.7 &
   -- & -- \\
 V709 Cas   & 0.96$\pm$0.05 & 30.0$\pm$2.5  & 0.88$\pm$0.06 &23.3$\pm$2.2 &
   0.95$^{+0.07}_{-0.08}$ & 25.6$^{+2.7}_{-2.4}$ \\
 NY Lup     & 1.12$\pm$0.06 & 39.1$\pm$4.3  & 0.99$\pm$0.06 &27.1$\pm$2.2 &
   -- & -- \\
 V2400 Oph  & 0.81$\pm$0.10 & 21.1$\pm$2.8  & 0.73$\pm$0.05 &18.6$\pm$1.4 &
    -- & --\\
 GK Per     & 0.90$\pm$0.12 & 26.2$\pm$5.4  & 1.00$\pm$0.40 &28.7$\pm$15.6 &
    0.87$^{+0.25}_{-0.21}$ & 23.0$^{+9.2}_{-6.5}$ \\
 V1223 Sgr  & 0.65$\pm$0.04 & 17.0$\pm$1.3  & 0.74$\pm$0.07 &18.8$\pm$1.2 &
    -- & --\\
 RX J2133   & 0.91$\pm$0.06 & 27.1$\pm$2.9  & 0.90$\pm$0.12 &23.8$\pm$4.3 &
    -- & --\\
 IGR J00234+6141 & 0.85$\pm$0.39 & 22.8$^{+30.0}_{-11.5}$ & 0.64$\pm$0.18 &15.9$\pm$5.1 &
    -- & --\\
 IGR J17303-0601 & 1.08$\pm$0.07 & 37.1$\pm$4.4 & 0.98$\pm$0.14 &26.7$\pm$4.8&
    1.10$^{+0.21}_{-0.16}$ & 31.6$^{+12.7}_{-7.8}$\\
 \noalign{\smallskip}
 \hline 
 \end{tabular}
 \label{tbl:ibis}
\end{table*}

\subsubsection{Comparison with INTEGRAL/IBIS data}

\cite{barlow06} and very recently \cite{landi08} provided a table with 
temperatures derived from Integral/IBIS data.
The method of Aizu (1973) provides a direct conversion from
the bremsstrahlung temperature to mass of the white dwarf.
which we use to derive the masses as listed in Table \ref{tbl:ibis}.
The comparison of these masses to those derived here using {\em Swift}/BAT
shows, in  general, very good agreement 
(Table \ref{tbl:ibis}. The only outlier is MU Cam where our temperature
(and mass) is substantially greater. However, in a recent analysis of
XMM-{\it Newton} and INTEGRAL data, \cite{staude08}
conclude that the estimate
of \cite{barlow06} seems  too low. These authors attempted a
combined XMM-INTEGRAL fit, and arrive at a temperature of 35$\pm$10 keV,
which they describe as ``convincing agreement with the INTEGRAL data points''.
Though the {\em Swift}/BAT spectrum for MU Cam is the worst in terms of
reduced $\chi^2$, we are confident that our measurement does not 
suffer from large systematic uncertainties.

%______________________________________________________________
\section{Conclusions}

An accurate determination of white dwarf masses in IPs is 
important for our understanding of the accretion geometry
and  binary evolution. However, this has so far been complex
because  of the difficulty in measuring the radial velocities
and inclination of the systems.
The main advantage of using X-ray measurements is that
they  allow a direct measurement of the WD mass without relying
on the above parameters.

Indeed, since bremsstrahlung is the main cooling mechanism in the post-shock
region, it is possible to relate the  temperature
of the IP's X-ray spectrum  to the WD mass \citep[e.g.][]{aizu73,suleimanov05}.
Using this method, we determined WD masses for a complete sample of 22
IPs detected in the ongoing {\it Swift}/BAT survey.
For 6 objects, this represents the only available mass estimate.

All mass estimates are in good agreement with previously estimated
WD masses and in general, thanks to the good signal-to-noise ratio, are more 
accurate. In particular, we reach surprisingly good agreement for the only
eclipsing IP, XY Ari, suggesting that our analysis does not suffer 
from major systematic errors.

Detailed analysis of the light curve of GK Per shows that BAT detects
the dependence of the source flux on the accretion rate.
While this has been expected for a long time, to our knowledge this effect 
has never yet been supported by observational evidence.

\begin{acknowledgements}
We are grateful to Valery Suleimanov for providing his model and for his
help.  JB acknowledges  extensive discussions with H. Ritter
and M. Revnivtsev. MA  acknowledges funding from the DFG Leibniz-Prize 
to Prof. G. Hasinger (HA 1850/28-1). We thank the referee for the quick
response. This research has made use of the NASA/IPAC 
extragalactic Database (NED) which
is operated by the Jet Propulsion Laboratory, of data obtained from the 
High Energy Astrophysics Science Archive Research Center (HEASARC) provided 
by NASA's Goddard Space Flight Center and of the SIMBAD Astronomical Database.
\end{acknowledgements}

\bibliographystyle{aa}
\bibliography{Brunschweiger_IP}

\begin{thebibliography}{48}
\expandafter\ifx\csname natexlab\endcsname\relax\def\natexlab#1{#1}\fi

\bibitem[{{Aizu}(1973)}]{aizu73}
{Aizu}, K. 1973, Progress of Theoretical Physics, 49, 1184

\bibitem[{{Ajello} {et~al.}(2008{\natexlab{a}}){Ajello}, {Greiner}, {Kanbach},
  {Rau}, {Strong}, \& {Kennea}}]{marco7a}
{Ajello}, M., {Greiner}, J., {Kanbach}, G., {et~al.} 2008{\natexlab{a}}, \apj,
  678, 102

\bibitem[{{Ajello} {et~al.}(2008{\natexlab{b}}){Ajello}, {Rau}, {Greiner},
  {Kanbach}, {Salvato}, {Strong}, {Barthelmy}, {Gehrels}, {Markwardt}, \&
  {Tueller}}]{marco7b}
{Ajello}, M., {Rau}, A., {Greiner}, J., {et~al.} 2008{\natexlab{b}}, \apj, 673,
  96

\bibitem[{{Allan} {et~al.}(1996){Allan}, {Hellier}, \& {Ramseyer}}]{Allan96}
{Allan}, A., {Hellier}, C., \& {Ramseyer}, T.~F. 1996, \mnras, 282, 699

\bibitem[{{Barlow} {et~al.}(2006){Barlow}, {Knigge}, {Bird}, {J Dean}, {Clark},
  {Hill}, {Molina}, \& {Sguera}}]{barlow06}
{Barlow}, E.~J., {Knigge}, C., {Bird}, A.~J., {et~al.} 2006, \mnras, 372, 224

\bibitem[{{Barthelmy}(2004)}]{barthelmy04}
{Barthelmy}, S.~D. 2004, in SPIE Conf., Vol. 5165, X-Ray and Gamma-Ray
  Instrumentation for Astronomy XIII. Edited by Flanagan, Kathryn A.; Siegmund,
  Oswald H. W. Proceedings of the SPIE, Volume 5165, pp. 175-189 (2004)., ed.
  K.~A. {Flanagan} \& O.~H.~W. {Siegmund}, 175--189

\bibitem[{{Berriman}(1987)}]{berriman87}
{Berriman}, G. 1987, \aaps, 68, 41

\bibitem[{{Beuermann} {et~al.}(2003){Beuermann}, {Harrison}, {McArthur},
  {Benedict}, \& {G{\"a}nsicke}}]{beuermann03}
{Beuermann}, K., {Harrison}, T.~E., {McArthur}, B.~E., {Benedict}, G.~F., \&
  {G{\"a}nsicke}, B.~T. 2003, \aap, 412, 821

\bibitem[{{Beuermann} {et~al.}(2004){Beuermann}, {Harrison}, {McArthur},
  {Benedict}, \& {G{\"a}nsicke}}]{beuermann04}
{Beuermann}, K., {Harrison}, T.~E., {McArthur}, B.~E., {Benedict}, G.~F., \&
  {G{\"a}nsicke}, B.~T. 2004, \aap, 419, 291

\bibitem[{{Bonnet-Bidaud} {et~al.}(2006){Bonnet-Bidaud}, {Mouchet}, {de
  Martino}, {Silvotti}, \& {Motch}}]{Bonnet-Bidaud06}
{Bonnet-Bidaud}, J.~M., {Mouchet}, M., {de Martino}, D., {Silvotti}, R., \&
  {Motch}, C. 2006, \aap, 445, 1037

\bibitem[{{Brat} {et~al.}(2006){Brat}, {Hudec}, {Simon}, {Strobl}, {Kubanek},
  {Nekola}, \& {Jelinek}}]{brat06}
{Brat}, L., {Hudec}, R., {Simon}, V., {et~al.} 2006, The Astronomer's Telegram,
  965, 1

\bibitem[{{Buckley} \& {Tuohy}(1989)}]{buckley89}
{Buckley}, D.~A.~H. \& {Tuohy}, I.~R. 1989, \apj, 344, 376

\bibitem[{{Crampton} {et~al.}(1986){Crampton}, {Fisher}, \&
  {Cowley}}]{crampton86}
{Crampton}, D., {Fisher}, W.~A., \& {Cowley}, A.~P. 1986, \apj, 300, 788

\bibitem[{{Cropper} {et~al.}(1998){Cropper}, {Ramsay}, \& {Wu}}]{cropper98}
{Cropper}, M., {Ramsay}, G., \& {Wu}, K. 1998, \mnras, 293, 222

\bibitem[{{Cropper} {et~al.}(1999){Cropper}, {Wu}, {Ramsay}, \&
  {Kocabiyik}}]{cropper99}
{Cropper}, M., {Wu}, K., {Ramsay}, G., \& {Kocabiyik}, A. 1999, \mnras, 306,
  684

\bibitem[{{de Martino} {et~al.}(2006){de Martino}, {Bonnet-Bidaud}, {Mouchet},
  {G{\"a}nsicke}, {Haberl}, \& {Motch}}]{demartino06}
{de Martino}, D., {Bonnet-Bidaud}, J.-M., {Mouchet}, M., {et~al.} 2006, \aap,
  449, 1151

\bibitem[{{de Martino} {et~al.}(2004){de Martino}, {Matt}, {Belloni},
  {Chiappetti}, {Haberl}, \& {Mukai}}]{deMartino2004}
{de Martino}, D., {Matt}, G., {Belloni}, T., {et~al.} 2004, Nuclear Physics B
  Proceedings Supplements, 132, 693

\bibitem[{{Frank} {et~al.}(2002){Frank}, {King}, \& {Raine}}]{Frank02}
{Frank}, J., {King}, A., \& {Raine}, D.~J. 2002, {Accretion Power in
  Astrophysics: Third Edition} (Accretion Power in Astrophysics, by Juhan Frank
  and Andrew King and Derek Raine, pp.~398.~ISBN 0521620538.~Cambridge, UK:
  Cambridge University Press, February 2002.)

\bibitem[{{Haswell} {et~al.}(1997){Haswell}, {Patterson}, {Thorstensen},
  {Hellier}, \& {Skillman}}]{haswell97}
{Haswell}, C.~A., {Patterson}, J., {Thorstensen}, J.~R., {Hellier}, C., \&
  {Skillman}, D.~R. 1997, \apj, 476, 847

\bibitem[{{Hellier}(1993)}]{hellier93}
{Hellier}, C. 1993, \mnras, 264, 132

\bibitem[{{Hellier}(1997)}]{hellier97}
{Hellier}, C. 1997, \mnras, 291, 71

\bibitem[{{Hellier}(2002)}]{hellier2002}
{Hellier}, C. 2002, in Astronomical Society of the Pacific Conference Series,
  Vol. 261, The Physics of Cataclysmic Variables and Related Objects, ed. B.~T.
  {G{\"a}nsicke}, K.~{Beuermann}, \& K.~{Reinsch}, 92--+

\bibitem[{{Hellier} {et~al.}(1991){Hellier}, {Cropper}, \& {Mason}}]{hellier91}
{Hellier}, C., {Cropper}, M., \& {Mason}, K.~O. 1991, \mnras, 248, 233

\bibitem[{{Hoogerwerf} {et~al.}(2004){Hoogerwerf}, {Brickhouse}, \&
  {Mauche}}]{hoogerwerf04}
{Hoogerwerf}, R., {Brickhouse}, N.~S., \& {Mauche}, C.~W. 2004, \apj, 610, 411

\bibitem[{{Ishida}(1991)}]{ishida91}
{Ishida}, M. 1991, {PhD thesis} (Univ. Tokyo)

\bibitem[{{King} \& {Lasota}(1979)}]{KingLasota79}
{King}, A. \& {Lasota}, J. 1979, \mnras, 188, 653

\bibitem[{{Landi} {et~al.}(2009){Landi}, {Bassani}, {Dean}, {Bird}, {Fiocchi},
  {Bazzano}, {Nousek}, \& {Osborne}}]{landi08}
{Landi}, R., {Bassani}, L., {Dean}, A.~J., {et~al.} 2009, \mnras, 392, 630

\bibitem[{{Littlefair} {et~al.}(2001){Littlefair}, {Dhillon}, \&
  {Marsh}}]{littlefair01}
{Littlefair}, S.~P., {Dhillon}, V.~S., \& {Marsh}, T.~R. 2001, \mnras, 327, 669

\bibitem[{{Markwardt} {et~al.}(2005){Markwardt}, {Tueller}, {Skinner},
  {Gehrels}, {Barthelmy}, \& {Mushotzky}}]{markwardt05}
{Markwardt}, C.~B., {Tueller}, J., {Skinner}, G.~K., {et~al.} 2005, \apjl, 633,
  L77

\bibitem[{{Mateo} {et~al.}(1991){Mateo}, {Szkody}, \& {Garnavich}}]{mateo91}
{Mateo}, M., {Szkody}, P., \& {Garnavich}, P. 1991, \apj, 370, 370

\bibitem[{{McArthur} {et~al.}(2001){McArthur}, {Benedict}, {Lee}, {van Altena},
  {Slesnick}, {Rhee}, {Patterson}, {Fredrick}, {Harrison}, {Spiesman}, {Nelan},
  {Duncombe}, {Hemenway}, {Jefferys}, {Shelus}, {Franz}, \&
  {Wasserman}}]{mcarthur01}
{McArthur}, B.~E., {Benedict}, G.~F., {Lee}, J., {et~al.} 2001, \apj, 560, 907

\bibitem[{{McHardy} {et~al.}(1987){McHardy}, {Pye}, {Fairall}, \&
  {Menzies}}]{mchardy87}
{McHardy}, I.~M., {Pye}, J.~P., {Fairall}, A.~P., \& {Menzies}, J.~W. 1987,
  \mnras, 225, 355

\bibitem[{{Morales-Rueda} {et~al.}(2002){Morales-Rueda}, {Still}, {Roche},
  {Wood}, \& {Lockley}}]{morales02}
{Morales-Rueda}, L., {Still}, M.~D., {Roche}, P., {Wood}, J.~H., \& {Lockley},
  J.~J. 2002, \mnras, 329, 597

\bibitem[{{Penning}(1985)}]{penning85}
{Penning}, W.~R. 1985, \apj, 289, 300

\bibitem[{{Ramsay}(2000)}]{ramsay00}
{Ramsay}, G. 2000, \mnras, 314, 403

\bibitem[{{Ramsay} {et~al.}(1998){Ramsay}, {Cropper}, {Hellier}, \&
  {Wu}}]{ramsay98}
{Ramsay}, G., {Cropper}, M., {Hellier}, C., \& {Wu}, K. 1998, \mnras, 297, 1269

\bibitem[{{Revnivtsev} {et~al.}(2004){Revnivtsev}, {Sazonov}, {Jahoda}, \&
  {Gilfanov}}]{revnivtsev04}
{Revnivtsev}, M., {Sazonov}, S., {Jahoda}, K., \& {Gilfanov}, M. 2004, \aap,
  418, 927

\bibitem[{{Rothschild} {et~al.}(1981){Rothschild}, {Gruber}, \&
  {Knight}}]{rothschild81}
{Rothschild}, R., {Gruber}, D., \& {Knight}, F. e.~a. 1981, \apj, 250, 723

\bibitem[{{Sabbadin} \& {Bianchini}(1983)}]{sabbadin83}
{Sabbadin}, F. \& {Bianchini}, A. 1983, \aaps, 54, 393

\bibitem[{{Staude} {et~al.}(2008){Staude}, {Schwope}, {Schwarz}, {Vogel},
  {Krumpe}, \& {Nebot Gomez-Moran}}]{staude08}
{Staude}, A., {Schwope}, A.~D., {Schwarz}, R., {et~al.} 2008, \aap, 486, 899

\bibitem[{{Suleimanov} {et~al.}(2005){Suleimanov}, {Revnivtsev}, \&
  {Ritter}}]{suleimanov05}
{Suleimanov}, V., {Revnivtsev}, M., \& {Ritter}, H. 2005, \aap, 435, 191

\bibitem[{{Sutherland} \& {Dopita}(1993)}]{sutherland93}
{Sutherland}, R.~S. \& {Dopita}, M.~A. 1993, \apjs, 88, 253

\bibitem[{{Szkody} \& {Silber}(1996)}]{szkody96}
{Szkody}, P. \& {Silber}, A. 1996, \aj, 112, 289

\bibitem[{{{\v S}imon}(2002)}]{simon2002}
{{\v S}imon}, V. 2002, \aap, 382, 910

\bibitem[{{Warner}(1987)}]{warner87}
{Warner}, B. 1987, \mnras, 227, 23

\bibitem[{{Warner}(1995)}]{warner95}
{Warner}, B. 1995, {Cataclysmic variable stars} (Cambridge Astrophysics Series,
  Cambridge, New York: Cambridge University Press, |c1995)

\bibitem[{{Watson} {et~al.}(1985){Watson}, {King}, \& {Osborne}}]{watson85}
{Watson}, M.~G., {King}, A.~R., \& {Osborne}, J. 1985, \mnras, 212, 917

\bibitem[{{Woelk} \& {Beuermann}(1996)}]{WoelkBeuer96}
{Woelk}, U. \& {Beuermann}, K. 1996, \aap, 306, 232

\end{thebibliography}

\end{document}